**Quantitative imaging of 478-keV prompt gamma rays from boron neutron capture reactions**

Tetsuya Mizumoto[a,b,*], Shotaro Komura[a,b,*], Atsushi Takada[b], Yoshinori Sakurai[c], and Toru Tanimori[b,c]

[a] Development department, J-BEAM, Inc., 1-7, Nakamaru, Yamadaoka, Naraha-machi, Futaba-gun, Fukushima 979-0513, Japan

[b] Department of Science, Kyoto University, Kitashirakawa Oiwake-cho, Sakyo-ku, Kyoto-shi, Kyoto 606-8502, Japan

[c] Institute for Integrated Radiation and Nuclear Science, Kyoto University, 2, Asashiro-nishi, Kumatori-cho, Sennan-gun, Osaka 590-0494, Japan

[*] Corresponding author: mizumoto@cr.scphys.kyoto-u.ac.jp, komura@cr.scphys.kyoto-u.ac.jp

**ABSTRACT**

The accurate imaging and quantitative measurement of 478-keV prompt gamma rays are critical for advancing boron neutron capture therapy (BNCT), a promising cancer treatment. Although numerical simulations have indicated that such measurements are feasible, their practical application has proven challenging. This study introduces a gamma-ray imaging detector designed specifically for precise BNCT measurements. Using boron-rich phantom samples, we successfully imaged 478-keV gamma rays and established a linear correlation between gamma-ray production and boron concentration. Furthermore, applying this technique in a recognized BNCT treatment facility demonstrated the detector's effectiveness in monitoring boron dose distribution during neutron irradiation, both in pre-treatment diagnostics and throughout the treatment process.

**INTRODUCTION**

Boron neutron capture therapy (BNCT) is a promising cancer treatment that targets difficult-to-treat tumors, especially those in the brain, head, and neck. This therapy relies on the neutron-boron reaction but faces challenges with respect to measuring treatment effectiveness. In BNCT, a nontoxic boron compound is administered to the patient. This compound selectively accumulates in tumor cells that are subsequently irradiated with a neutron beam. The interaction between neutrons and the $^{10}$B (boron) within the tumor produces high-energy alpha particles and $^{7}$Li nuclei. Owing to their high linear energy transfer and short range (~10 μm), these particles can effectively destroy cancer cells in boron-rich tumors. Notably, one of the key requirements for boron compounds used in BNCT treatment is high tumor uptake ($^{10}$B concentration of 20 ppm or more) and low uptake in normal tissues, with tumor-to-normal tissue and tumor-to-blood boron concentration ratios of at least 3:1[1]. To ensure effective treatment while protecting healthy tissue, monitoring the $^{10}$B (n, a) $^{7}$Li reaction is crucial. This is typically achieved by detecting 478-keV prompt gamma rays, which are produced in 94% of neutron-boron reactions. Moreover, the measurement of 2223-keV prompt gamma rays, resulting from the interaction between thermal neutrons and $^{1}$H atoms, distributed throughout the body, provides data on boron concentration and thermal neutron fluence in the body.

Several methods have been proposed for developing fast gamma-ray imaging devices for detecting 478-keV gamma rays. Kobayashi et al. introduced the prompt gamma-ray single-photon emission computed tomography (PG-SPECT) technique, combining prompt gamma-ray analysis with SPECT to enable noninvasive and three-dimensional (3D) dose estimation during BNCT[2-6]. The PG-SPECT system uses heavy metal collimators and gamma-ray detectors with high energy resolution. To effectively measure 478-keV gamma rays, a thick collimator is required to block rays from penetrating the collimator material. However, this reduces sensitivity compared with conventional SPECT. Although the thick collimator minimizes gamma-ray loss, it introduces noise from 511-keV annihilation and scattered gamma rays. Moreover, the fixed direction of neutrons in PG-SPECT limits the rotation of the target, thus affecting the accuracy of 3D imaging.

In theory, conventional Compton cameras can simultaneously detect 478-keV and 2223-keV prompt gamma rays generated from the $^1$H neutron capture reaction[7]. Several Monte Carlo simulation studies have examined the feasibility and design of these cameras as gamma-ray detectors[8-10]. Although Compton cameras offer benefits such as a broad energy range, a wide field of view, and the ability to operate without the presence of thick collimators, noise reduction remains a major issue, especially because of accidental coincidence events. These cameras require simultaneous signals from the scatterer and absorber but struggle to fully resolve the Compton equation, thus making it difficult to distinguish real signals from accidental coincidences at high count rates. Moreover, reconstructing incident gamma rays in a single direction without information on the electron recoil direction is impossible as it results in a conical surface centered at the Compton scattering point, potentially leading to overlapping events and complicating quantitative imaging[11]. Recent experiments, such as Sakai et al.'s 2023 study, used a high-energy resolution Si/CdTe Compton camera to successfully image 478-keV gamma rays from boron neutron capture reactions[12], revealing well-defined energy peaks and qualitative visualizations. In this experiment, a $^{241}$Am neutron source was used in the environment with a neutron flux that was much lower than that encountered in BNCT.

To tackle the quantification challenge in BNCT, electron-tracking Compton cameras (ETCCs) have been developed. These cameras can precisely measure the direction of Compton recoil electrons using tracking data, allowing event-by-event reconstruction of sub-MeV/MeV gamma rays. Unlike conventional Compton cameras, ETCCs function as fully bijective gamma cameras based on optical principles, maintain a linear relationship between image brightness and gamma-ray intensity[13]. This results in high-contrast images with minimal interference from surrounding areas. Although conventional Compton cameras generate nonlinear images, ETCCs produce linear gamma-ray images with a wide field of view (>1 sr). ETCCs measure all parameters of the Compton scattering process, such as the energy loss rate (dE/dx) of the recoil electron and the angle ($\alpha$) between the gamma-ray scattering direction and the electron's recoil direction, thereby fully solving the Compton equation. This capability allows ETCCs to enhance noise reduction and efficiently differentiate real signals from background noise. An ETCC comprises two subdetectors: a time projection chamber (TPC), scattering the incident gamma ray and tracks the recoil electron, and a scintillation camera, absorbing the scattered gamma ray. Various ETCCs have been developed for numerous applications such as MeV gamma-ray astronomy[11,13-14], environmental gamma-ray measurements[15-17], and medical applications[18-20].

In our previous study, we developed a compact ETCC using gadolinium orthosilicate ($Gd_2SiO_5$:Ce; GSO) scintillators (hereafter referred to as GSO-ETCC), featuring a 20-cm TPC to enhance the efficiency of gamma-ray detection[21]. Despite its large-volume TPC, GSO-ETCC is compact, similar in size to the existing 10-cm TPC systems. It includes a scintillation camera with updated readout circuits, pixel scintillator arrays (PSAs) of GSO crystals, multi-anode photomultiplier tubes (PMTs), and a new high-speed data acquisition system using Ethernet data transfer. To further improve energy and imaging resolution, we developed an advanced ETCC(referred to as GAGG-ETCC) using multi-pixel photon counters (MPPCs) and PSAs comprising a high-energy-resolution gadolinium aluminum gallium garnet ($Gd_3(Al,Ga)_5O_{12}(Ce)$; HR-GAGG) crystals in the scintillation camera. Using HR-GAGG in place of GSO improved the 511 keV gamma-ray energy resolution [full width half maximum (FWHM)] of ETCCs from 12.7% ± 1.3% to 6.6% ± 0.4%, resulting in better separation of the 478-keV peak from continuous component and 511-keV annihilation line gamma rays[16]. The ETCC provides two key angular resolution parameters related to the reconstruction of the incident gamma-ray direction, i.e., the angular resolution measure (ARM) and scattering plane deviation (SPD). The ARM is defined as the difference between the Compton scattering angle calculated from energy measurements and the angle derived geometrically from the detected positions of the recoil electron and scattered gamma ray. This parameter quantifies the accuracy of the reconstructed Compton scattering angle. The other angular resolution parameter, i.e., SPD, is defined as the angular deviation of the reconstructed scattering plane from the real scattering plane, determined by the vectors of the incoming and scattered gamma rays. SPD represents the accuracy of reconstruction of the azimuthal scattering angle. Based on the measurements of the $^{22}$Na point source at the center of the field of view, the ARM and SPD distributions (FWHM) of GAGG-ETCC for 511-keV gamma rays were 5.7° ± 0.3° and 217° ± 62°, respectively. The measurement of $^{137}$Cs placed at the center of the field of view and at a distance of 244 mm from the detection area of the TPC exhibited a detection efficiency of $(2.04 ± 0.05) × 10^{-5}$ for the 662-keV gamma radiation of the GAGG-ETCC. For this measurement, a 3-mm-thick acrylic sheet and a 0.5-mm-thick lead sheet were placed between the TPC and the $^{137}$Cs source. Notably, the uncertainty in the detection efficiency includes the statistical error and certified accuracy of the radioactivity of the source. To validate the experimental results, a Geant4 simulation was performed using the same geometric configuration as that in the experiment[22,23]. The simulation yielded a detection efficiency of $(2.13 ± 0.03) × 10^{-5}$, corresponding with the experimental result within a value that is twice that of the experimental uncertainty. In the case of collimated 478- and 662-keV gamma rays from the viewing direction, the simulated detection efficiencies were $(1.224 ± 0.003) × 10^{-4}$ and $(5.89 ± 0.02) × 10^{-5}$, respectively. An overview of GAGG-ETCC can be found in the subsection "The ETCC apparatus" of the Methods section, with additional details available elsewhere.

This study demonstrates the quantitative imaging of 478-keV prompt gamma rays from boron neutron capture reactions in boron solution water phantoms using GAGG-ETCC. Experiments were performed at the Kyoto University Research Reactor (KUR) of the Institute for Integrated Radiation and Nuclear Science, Kyoto University (KURNS), with a thermal power setting of 1 MW. To test the basic quantitative imaging capabilities of 478-keV gamma rays, we conducted experiments at the E-3 neutron guide tube thermal neutron beamline at KUR. A thin boron solution phantom with a 4.5 mm diameter was placed near the ETCC and exposed to a low-thermal neutron beam at a data acquisition rate of ~30 Hz. Furthermore, to replicate real-world BNCT treatment conditions, further experiments were conducted at the heavy water neutron irradiation facility (KUR-HWNIF). Our experiments effectively demonstrated the capability of GAGG-ETCC to quantitatively monitor and image 478-keV prompt gamma rays from a boron-containing cubic phantom irradiated with neutrons from a distance greater than 2.5 m from the ETCC.

RESULTS

***Quantitative measurement of 478-keV prompt gamma rays from a small phantom placed at a close range***

An imaging test was performed on 478-keV gamma rays using a thin boric acid ($H_3BO_3$) solution phantom at the KUR E-3 thermal neutron beamline with ETCC. The neutron beam has a rectangular cross-section of 10 mm by 74 mm[24]. According to KURNS, the thermal neutron flux at the E-3 neutron guide's output port is $4 \times 10^5$ n/cm$^2$/sec at 1 MW.

To reduce neutron interactions with the container, perfluoroalkoxy alkane (PFA) cylindrical containers, which do not contain hydrogen, were used for the phantom solution, as illustrated in **Fig. 1a**. Boric acid solutions with varying $^{10}$B concentrations were prepared. The experimental setup is shown in **Fig. 1b** and **1c**. The phantom solution was positioned directly above ETCC, aligned with the E-3 tube neutron irradiation. The distance between the top most surface of the detection area of TPC and the center axis of the neutron beam is approximately 12 cm. The coordinates and geometrical details are summarized in **Fig. 1c**. We measured 478-keV prompt gamma rays from neutron irradiation of boron solutions with concentrations of 0, 462, 926, and 1385 ppm for several hours at a fixed position relative to ETCC. **Table 1** summarizes the experimental parameters. The signal and data acquisition rate showed no remarkable differences across the $^{10}$B concentrations, suggesting that the signal was primarily owing to background gamma rays generated within the laboratory.

**Figure 2** displays the energy spectra and 478-keV gamma-ray images for different $^{10}$B concentrations. The red energy spectrum in the top panel of each figure shows events remaining after applying the dE/dx cut, the TPC fiducial volume cut, selection of forward gamma-ray-like events, and an image cut with a 50-mm radius. The blue spectrum represents accidental coincidence events using the same sampling patterns as the red spectrum. The green spectrum in the middle panel is obtained by subtracting the blue spectrum from the red spectrum, representing the measured gamma-ray spectrum. All spectra clearly show a peak at 478 keV, with intensity proportionally increasing to the boron concentration (**Fig. 3a**, where all spectra are superimposed). Moreover, **Fig. 2a** shows a peak at 478 keV even with a 0-ppm boron concentration, which is speculated to be owing to prompt gamma rays from the $^{10}$B in the ETCC readout circuit board and interactions with scattered neutron rays. Using maximum likelihood-expectation maximization (ML-EM) for image reconstruction in the horizontal plane along the neutron beam's center axis, the bottom panel of each figure shows back-projected images for events in the energy range within 478 keV ± 5%. All gamma-ray images in this study were generated after 25 iterations of ML-EM. The color scale of all images is normalized so that the maximum of the image for **Fig. 2d** is 1. The ratio of the areas within a 50 mm radius from the phantom center in the images in**Fig. 2a-2d** was adjusted to match the ratio of the event rates within 478 keV ± 5% of the red spectrum at 478 keV in**Fig. 2a-2d**. For phantoms with concentrations of 462, 926, and 1385 ppm, the images converge to a single point indicating the location of the phantom, and the brightness increases with increasing concentration.By contrast, the 0-ppm phantom image does not exhibit this convergence. The spread of brightness is approximately 1 cm, accounting for the image resolution of ETCC and the uncertainty owing to the 5-cm vertical length of the phantom.

**Figure 3b-3e** illustrates superimposed gamma-ray–like spectra obtained after excluding accidental coincidences. Each subfigure corresponds to a boron solution phantom with a different $^{10}$B concentration, i.e., 0, 462, 926, and 1385 ppm. Spectra are shown for different image cut radii. These images show that as the image cut radius decreases, the phantom-derived 478-keV gamma-ray peak

component becomes more pronounced than the continuous component. This indicates the high imaging spectroscopy capability of ETCCs.

**Figure 4** illustrates the relationship between the measured detection rate of the 478-keV peak and the concentration of $^{10}$B in the solution. The detection rate was determined by fitting a function to the energy region from 380 to 580 keV in the gamma-like event spectrum with accidental coincidences excluded. The fitting function includes a linear term and two Gaussian distributions centered at 478 and 511 keV (refer to the "Fitting function of spectra" subsection of the "Methods" section). The function of the fitting results in the spectra. The fitting function of the spectra with image cuts of 50-mm radius for each concentration is illustrated in the green spectra of the middle panel in **Fig. 2**. The area of the Gaussian distribution centered at 478 keV over the entire energy range was defined as the detection rate of the 478-keV peak. In **Fig. 4**, the magenta and green data points represent the detection rate with and without a 50-mm radius image cut, respectively. These rates are compared with the result for a boron solution with a concentration of 1385 ppm. The blue line represents simulation results where the boron solution, container, acrylic plate, and air were modeled in the same positions as in the experiments. The simulation was conducted using the Particle and Heavy Ion Transport code System (PHITS), version 3.240[25]. The entire phantom was irradiated with thermal neutrons following a Maxwell-Boltzmann energy distribution at 293 K to match the experimental conditions. The flux of 478-keV prompt gamma rays passing through the TPC face was then estimated. The simulation results are normalized so that the detection rate at 1385 ppm $^{10}$B concentration is 1. The green data points show a discrepancy in the detection rate at 0 ppm $^{10}$B compared with the simulated results (blue line). However, this discrepancy is considerably reduced with the 50-mm image cut, as shown by the close alignment of the red data point with the blue line. These results demonstrate the excellent imaging and spectroscopic capabilities of ETCC.

*Measurement under high-intensity BNCT*

This study was conducted at the KUR heavy water neutron irradiation facility (KUR-HWNIF)[26], which is known for its extensive history of BNCT treatments. KUR-HWNIF supports research in medical biology, chemistry, pharmacy, and physical engineering and offers three irradiation modes with varying neutron energy distributions: thermal, epithermal, and mixed mode. For this study, the epithermal neutron irradiation mode was used at 1 MW operation, with thermal, epithermal, and fast neutron fluxes of $6.1 \times 10^6$, $1.5 \times 10^8$, and $9.5 \times 10^6$ n/cm$^2$/sec, respectively, at the reference point[27].

**Figure 5** shows the layout of the measurement setup. The phantom, comprising a 10-cm cubic acrylic container filled with the solutions of varying $^{10}$B concentrations, is placed on a LiF-polyethylene block, which is then set on a boron-free carrier. A neutron collimator wall with a 30 cm × 30 cm square aperture is positioned in front of the carrier. The walls, ceiling, and floor of the measurement room are lined with boron-containing polyethylene sheets. The ceiling made of 1-m-thick concrete has two vertically drilled cylindrical apertures located at different distances from the reactor core. ETCC is mounted facing downward on the roof, aligning with the center of the nearest aperture to the reactor. This aperture corresponds to the left experimental tunnel shown in Fig. 2 inreference [26]. At the phantom's location, the neutron flux is approximately $1 \times 10^7$ n/cm$^2$/sec during epithermal neutron irradiation at 1 MW, owing to the ~1 m separation from the bismuth window[26]. Although the neutron production rate during measurement is only one-fifth of that in actual BNCT treatments, the neutron flux to the phantom is considerably lower owing to its increased distance from the bismuth window. This results in a small ratio of prompt gamma rays from the phantom compared with background noise from the reactor core and room, making the signal-to-noise ratio challenging. The accuracy of the ETCC and phantom's positioning relative to the ceiling hole, as well as the ETCC tilt, was confirmed using a ruler, a digital inclinometer, and a laser pointer. Furthermore, the position's reproducibility was verified by adjusting the dolly to close the shutter when the reactor was off. Consequently, the phantom placement accuracy was ± 4 mm, the ETCC tilt was ± 0.15 degrees, and the alignment of the phantom in the gamma-ray image was accurate to ± 9 mm. The x–y coordinates used for measurement are shown in**Fig. 5a** and **5b** and are similar to those in **Fig. 1c** from the ETCC perspective. Several measurements were taken using four $^{10}$B concentration solutions (0, 46, 93, and 230 ppm). For each measurement, the phantom's center was offset from the coordinate origin along the x-axis. **Table 2** summarizes the conditions for each measurement, including the signal and data acquisition rates. The TPC signal rate, scintillation camera signal rate, and ETCC data acquisition rate were in the range of a few kHz to several hundred Hz for all measurements using the epithermal neutron irradiation mode. The data acquisition system had a dead time of less than 3%, which is suitable for the required data acquisition speed.

We analyzed the results from the phantom measurements taken at the center position of (x = 20 mm, y = 0 mm) for 0-, 46-, and 93-ppm concentrations. These levels are similar to those typically found in tumors treated with boronophenylalanine. **Figure 6a–6c** presents the measured spectra and ML-EM-processed back-projection images of 478-keV gamma rays in the horizontal plane passing through the centers of the bismuth window of the KUR-HWNIF, neutron collimator, and phantom. The green spectrum in the central panel of each figure shows gamma-ray-like events after removing accidental coincidences. In **Fig. 6a**, the 0-ppm spectrum displays a peak at 511 keV but no peak at 478 keV. By contrast,**Fig. 6b** and **6c** shows that the 478-keV peak becomes more prominent as the boron concentration increases. The lower panel of each figure features back-projection images processed with the ML-EM algorithm for events in the energy range within 478 keV ± 5% (with the maximum color scale normalized to 1). The color scale of the ML-EM-processed images is normalized such that the maximum value of the image for **Fig. 6c** is 1. The ratios of the areas within a 500-mm radius from the phantom center in **Fig. 6a–6c** were adjusted to match the ratio of the event rates within 478 keV ± 5% of the red spectrum at 478 keV in **Fig. 6a–6c**. In the images for **Fig. 6a–6c**, the reactor core is positioned toward the lower part, i.e., along the negative y-axis direction, and the outline of the phantom is shown with a white border without filling. The images for 46 and 93 ppm highlight the brightest region within the phantom, which becomes increasingly bright with increasing concentration, reflecting the increased intensity of the 478-keV gamma radiation observed in the spectra.

Furthermore, we investigated the impact of 511-keV annihilation gamma rays by comparing two phantoms: one filled with water and the other with a 230-ppm $^{10}$B solution. The green spectrum in the middle panel of **Fig. 7a** represents gamma-ray like energy, excluding random coincidence events. The spectral analysis shows a prominent peak at 478 keV and a second peak at 511 keV. The left and right images in the bottom panel of **Fig. 7a** display ML-EM processed back-projection images for events in the energy range within 478 keV ± 5% and within 511 keV$^{+5\%}_{-0\%}$, respectively. The ML-EM-processed images in **Fig. 7** are the images obtained by back-projecting in the same horizontal plane as shown in **Fig. 6**. To reduce the interference of 478-keV prompt gamma rays in the 511-keV image (the right image in the bottom panel of **Fig. 7a**), the gamma-ray image was generated for an energy range slightly above 511 keV. The color scale of the two images in **Fig. 7a** is normalized such that the maximum value of the left image is 1. The ratio of the areas within a 500 mm radius from the 230-ppm phantom center in the left image to that in the right image was adjusted so that it matches the ratio of the event rates within 478 keV ± 5% of the red spectrum to twice the event rates within 511 keV$^{+5\%}_{-0\%}$ of the same spectrum. In the left image of the bottom panel of **Fig. 7a**, the bright region corresponds to the phantom with 230-ppm $^{10}$B concentration, thus verifying the accurate detection of 478-keV gamma rays. By contrast, the right image in the bottom panel of **Fig. 7a** shows that the bright regions for 511-keV gamma rays are evenly distributed around the centers of the two phantoms, rather than aligning with the areas of high boron concentration. The 511-keV gamma rays, comprising continuous and peak components from annihilation, are not influenced by the $^{10}$B distribution. This difference in distribution highlights the distinct origins of the 478-keV and 511-keV gamma rays.

**Figure 7b** presents measurements from a phantom filled with a $^{10}$B solution at 230-ppm, positioned 9-cm off-center in the field of view with the following coordinates: x, 90 mm; y, 0 mm. The green spectrum in the middle panel clearly shows the 478-keV peak above the 511-keV peak. The ML-EM processed image shows that the bright region aligns accurately with the phantom. This demonstrates that 478-keV gamma rays can be detected even when the phantom is considerably displaced from the center of the field of view, with the gamma-ray image accurately reflecting the emission location within a few centimeters.

**Figure 8** illustrates the relationship between the detection rate of 478-keV gamma rays and the concentration of $^{10}$B in the solution at KUR-HWNIF. The red and green markers in the figure represent the experimental data, whereas the blue line represents the simulation result obtained using PHITS for a phantom positioned at the coordinates of x = 20 mm and y = 0 mm. In the PHITS simulation, neutrons were isotropically emitted from a single point located 1 meter toward the reactor core from the position directly beneath the ceiling hole of the irradiation room, near the bismuth window. The phantom and concrete ceiling with the hole were modeled, while the other regions were assumed to be filled with air. The simulation estimated the flux of 478-keV gamma rays entering ETCC through the concrete hole. The energy distribution of the emitted neutrons was determined according to the spectrum for the CO-0000-F epi-thermal neutron irradiation mode, as shown in Fig. 14(c) in reference [26], presenting multiple spectra. The measured and simulated results were normalized to a concentration of 93 ppm. The comparison includes data for the concentration of 230 ppm, which corresponds to the measurement number 4, and various phantom locations from measurements 1–3. Positional effects were adjusted by calculating the ratio of simulation results for the 230-ppm phantom under the same conditions as measurements 1–3 to those of measurement number 4. **Figure 8** reveals a strong correlation between the measured and simulated results, which are in agreement within the experimental uncertainties. Comparison with **Fig. 4**, the concrete and LiF-polyethylene plate effectively shielded the neutrons, preventing any neutrons from entering ETCC. Furthermore, because no 478-keV prompt gamma rays from the surroundings, other than through the hole, reached ETCC, no considerable difference was observed in the relationship between gamma-ray detection efficiency and boron concentration, with or without the image cut.

**DISCUSSION**

We measured 478-keV prompt gamma rays from boron-containing phantoms irradiated with neutrons using ETCC at two different facilities. First, we irradiated a phantom placed directly in front of ETCC at the E-3 neutron irradiation facility. Second, we irradiated phantoms with ~100 ppm $^{10}$B at one-fifth of the neutron generation rate used in BNCT therapy at the KUR-HWNIF BNCT treatment facility. For this second setup, ETCC, located outside the irradiation room, measured for 478-keV gamma radiation. In both instances, we successfully visualized the boron neutron capture reaction and quantitatively assessed concentration differences in 478-keV gamma ray emissions. These results highlight the ETCC effective imaging spectroscopy capabilities. The following sections explore the potential applications and insights derived from these measurements.

At the E-3 irradiation facility, we measured $^{10}$B concentrations of 0, 462, 926, and 1385 ppm. The results demonstrated the effective imaging spectroscopy capabilities of ETCC, which reduced noise by integrating event selection with imaging techniques. We achieved an image resolution of ~1 cm at a distance of ~12 cm from the detector surface and confirmed a quantitative correlation between the $^{10}$B concentration and the detection rate of 478-keV gamma rays. Notably, based on previous case reports, boron accumulation in tumors is approximately several tens of ppm; however, herein, a phantom with a boron-10 concentration exceeding 462 ppm was used, which is considerably higher than the concentration in previous studies owing to the low neutron flux and the small beam and phantom sizes, the amount of 478-keV gamma rays generated from a phantom with a boron-10 concentration of several tens of ppm is extremely low. Therefore, to obtain quantitative results within the limited machine time, the boron-10 concentration in the phantom was increased. The verification of these results at boron concentrations corresponding to actual accumulation in tumors will need to be performed in future studies. According to the simulation, when neutrons were irradiated under the same reactor thermal power conditions, the ratio of the 478-keV prompt gamma ray incident rate on ETCC between the 10-cm cubic phantom with a $^{10}$B concentration of 10 ppm and the phantom used in E-3 with a $^{10}$B concentration of 462 ppm was 0.93, thus indicating comparable rates. The ratio of gamma-ray incident rates in the 0–3 MeV energy range was 9.1. However, because the measurements at KUR-HWNIF were performed with a signal rate >500 Hz, the measurements were considered feasible. Therefore, the quantitative imaging of 478-keV gamma rays could be performed with the ETCC under the conditions of 5-MW reactor power, using a cubic boron solution phantom with 10-cm sides, even with low $^{10}$B concentrations, such as 10 ppm, which is below the tumor accumulation concentration in BNCT. By contrast, imaging a boron solution phantom with a lower volume and a $^{10}$B concentration of 10 ppm requires developing a more efficient ETCC with a higher detection sensitivity.

To evaluate the ETCC potential for BNCT diagnosis in humans, we measured 478-keV prompt gamma rays from a boron-containing phantom in the BNCT treatment room. Despite the ~2.5 m distance between the ETCC and the phantom, we achieved imaging of 478-keV prompt gamma rays with a resolution less than 10 cm and confirmed the ability to distinguish between the 478-keV and 511-keV gamma ray images. Measurements at varying boron concentrations showed a linear relationship between the detection rate and generation rate of 478-keV prompt gamma rays or the incidence rate into ETCC. During the experiment, the neutron generation rate at KUR-HWNIF was one-fifth of the rate used in actual BNCT treatment; thus, the noise environment of this experiment was similar level. By contrast, the epithermal neutron flux near the phantom was ~$10^7$/cm$^2$/sec, which is about two orders of magnitude lower than BNCT treatments because of the long distance between the beam exit and phantom, resulting in a considerably lower signal-to-noise ratio than the actual clinical conditions. Thus, the usual set up of BNCT can improve the signal-to-noise ratio of this experiment dramatically.

The concrete ceiling in the KUR-HWNIF irradiation room acted as a protective barrier for ETCC, shielding it from neutrons and background gamma rays. By optimizing shielding and reducing the distance to the phantom, the spatial resolution of ETCC would be considerably improved. For example, reducing the concrete shielding to 50 cm and the distance to 30 cm could cut the measurement target distance by one-third, thereby substantially increasing the influx of 478-keV prompt gamma rays into ETCC. In addition, shielding ETCC from gamma rays from the reactor core would improve its performance, making it suitable for actual BNCT treatments with enhanced spatial resolution. Further improvements could be made by refining Compton scattering point calculations and determining the direction of recoil, which can be achieved by adjusting the pitch of electron track sampling in TPC from 800 to 400 μm and using deep learning for track analysis[28]. Furthermore, replacing the scintillation camera with CdZnTe (CZT) semiconductor detectors could greatly enhance the energy resolution of ETCC, allowing for spatial resolution measurements of less than 1 cm. However, as the count rate environment increases, accidental coincidence events within the background noise also increase, which may cause measurements with ETCC to be more challenging. To address this drawback, varied improvements approaches can be attempted, such as modifying the size and design of ETCC or enhancing the selectivity of 478-keV prompt gamma ray events from background noise, including accidental coincidences, through better energy resolution. Other approaches can include strengthening shielding to prevent the entry of neutrons and gamma rays located outside the region of interest and the development of advanced analysis methods. To verify the suitability of ETCC for BNCT treatment monitoring, it is essential to design and develop the ETCC, including its shielding component, in such a way that it is compatible with the treatment conditions. Further investigations are required to determine the practical feasibility of real-time monitoring through ETCC.

**METHODS**

*ETCC apparatus*

ETCC is an advanced Compton camera designed to measure the direction of Compton recoil electrons and produce detailed images. It comprises two main components: a time projection chamber (TPC) that monitors Compton scattering and a scintillation camera that detects the scattered gamma rays.

TPC is a gaseous detector with a cylindrical drift cage that is about 20-cm in height. It features a gas electron multiplier (GEM) that is a 100-μm thick and nearly circular, with a diameter of about 20 cm, and a regular octagonal micro-pixel chamber (μPIC) with 20-cm sides. The μPIC is a micro-pattern gaseous detector with anode and cathode electrodes arranged in orthogonal strips spaced 400 μm apart. The signals from adjacent electrodes are combined in the readout circuit, resulting in a pitch of 800 μm and a μPIC detection area of $\sim 3.3 \times 10^2$ cm$^2$. The GEM acts as an additional gas amplifier and is positioned parallel to μPIC. Inside the TPC drift cage, the electric field is uniform, with an applied voltage of ~200 V/cm between the top and bottom surfaces, leading to an electron drift velocity of ~4 cm/μs, which was determined from data on the drift region and the time width of muon and other particle tracks that penetrate both ends of the drift region. These components are housed in an aluminum pressure vessel filled with a gas mixture of Ar-CF$_4$-isoC$_4$H$_{10}$ in a ratio of 95:3:2. The electron cloud produced by charged particles moves steadily under the influence of the electric field. The amplified charge is collected from each strip electrode of μPIC. The three-dimensional (3D) track shape and position information are derived from the strip position and drift time. In addition, the 3D track position within the TPC is determined by measuring the time difference between the detection of scattered gamma rays in the scintillation camera and the TPC signal. The TPC readout boards process signals from 128 strips, recording digital track data, timestamps, and 32 strip-sum waveforms as detailed in reference [29]. Data from these boards are sent to a personal computer (PC) via Gigabit Ethernet and TCP using SiTCP technology, as described in reference [13].

The scintillation camera features pixel scintillator arrays (PSAs), MPPCs, and associated readout circuits and manages the ETCC trigger. It uses high-energy resolution $Gd_3(Al,Ga)_5O_{12}(Ce)$ (HR-GAGG) pixelated scintillation crystals (C&A company, Japan) for PSAs. Each pixel measures 6.1 mm × 6.1 mm × 15 mm, with four 4 × 4 pixel arrays making up one PSA. Each pixel is optically isolated using Vikuiti enhanced specular reflectors (3M, USA), and MPPC S13361-6050NE-04 devices from Hamamatsu Photonics Co. Ltd. serve as photodetectors. PSAs and MPPCs are coupled with optical grease (V-788, Adhesive Materials Group, USA) and housed in a light-shielded, precisely sized thin aluminum box. The scintillation camera includes a 6 × 6 array of PSAs, arranged with a pitch of 28 mm. The 16 output signals from each MPPC module are processed using a resistor matrix, producing four corner signals. These signals are then amplified and digitized using a 14-bit 100 MSPS ADC and analyzed by an FPGA. FPGA integrates the waveform over time to determine the magnitude of each output signal and provides information on the time stamp, pixel position where the signal was generated, and the sum integral value of the four corner outputs. The signal readout and data acquisition control unit of scintillation camera includes four digital input/output ports for the TPC readout boards. This unit handles TPC hit signals, data acquisition triggers, and timestamp calibration signals. It ensures synchronized data acquisition, capturing TPC and scintillation camera signals simultaneously within 8 μsec. The data from the scintillation camera, including TPC signal rates and operating parameters, is transferred to a PC via Gigabit Ethernet.

*ETCC data analysis method*

Data from the scintillation camera and the four TPC readout boards are recorded on tPC during measurements. Gamma-ray images or spectra are produced through offline data analysis, including calculating the energy, Compton scattering vertex, and recoil direction of Compton recoil electrons based on TPC data. The process begins by adjusting the positional data along the drift direction (z-axis) of the tracks taking, considering the time difference between timestamps and the known drift velocity of TPC. This provides two-dimensional (2D) track information
(x-z and y-z) from the anode and cathode readout boards of μPIC. The electron track analysis method, as described in reference [30], is then applied to determine the Compton scattering vertex and recoil direction. Finally, the recoil electron energy is calculated by integrating the FADC data waveform signal and using the calibration functions for each 8 × 8 µPIC area.

The absorption position is identified using the scintillation camera data by referring to a table that matches pixel IDs with positions. Absorption energy is calculated using the calibration function for each pixel. By integrating data from the TPC and the scintillation camera, we reconstruct the incident gamma radiation direction and energy. However, the ETCC output provides accurate gamma-ray data and includes noise events such as coincidental TPC occurrences, electrons from other gamma rays, and unrelated charged particles like muons and scattered atomic nuclei. In addition, some gamma-ray events may be missed if electrons escape the detection area, leading to incomplete reconstruction of the gamma ray direction or energy. To filter out noise events and focus on specific gamma-ray incidents, the following criteria were used:

1. **dE/dx cut:** This criterion removes minimum ionization particle (MIP)-like events by isolating fully contained electron-like events in a 2D histogram that plots track range against deposition energy in TPC.

2. **TPC fiducial volume cut:** This criterion eliminates events with tracks detected within 5 mm of the µPIC edges, thereby removing electrons that enter from outside the effective detection volume.

3. **Selection of forward gamma-ray-like events:** This criterion selects events where the reconstructed gamma-ray direction is forward.

4. **ETCC energy cut:** This energy filter identifies gamma-ray events in a narrow energy range, usually within 5% of 478 keV.

5. **Image area cut:** To refine the data, a back-projected image of the gamma rays was created using

the previous selection criteria. Events were then selected from a specific region of the image. This selection considered the two angular accuracies of ETCC for the reconstructed gamma-ray direction: the image blur is based on the distributions of ARM and SPD. ARM follows a Cauchy distribution with a FWHM of 5°, whereas SPD follows a Gaussian distribution with an FWHM of 250°. Herein, the "no image cut" spectrum includes events across the entire back-projection plane, whereas the "image cut" spectrum focuses on a specific region with a 50-mm radius for the E-3 experimental results and 500 mm for the KUR-HWNIF results.

The application of criteria 1–4 generated gamma-ray images, whereas conditions 1–3 and 5 generated energy spectra. However, these cuts do not entirely eliminate accidental coincidences. To address this issue, the gamma-ray spectrum containing accidental coincidence events was subtracted from the spectrum of coincidence events. The accidental coincidence spectrum was obtained by analyzing events where the z-coordinate of the proposed Compton scattering point fell outside the TPC drift region. If the z-coordinate is above the drift top, it indicates a stochastic event not related to a single gamma ray. Events with z-coordinates between 26.5 and 66.5 mm above the drift top are considered accidental coincidences, and pseudo-event coincidences are generated by randomly reallocating the z-coordinates of the candidate Compton scattering points within the drift region. As the z-coordinate distribution for accidental coincidences is consistent, whether inside or outside the drift region, the ratio of accidental to non-accidental coincidences can be determined. This ratio was then used to weight the accidental coincidence events when creating the spectra of the coincident events.

*Fitting function of spectra*

We use a complex function that combines two Gaussian distributions and a linear function:

$$f(x) = c_1 \cdot e^{-\frac{(x - c_2)^2}{2c_3^2}} + c_4 \cdot e^{-\frac{\left(x - \frac{511}{478} \cdot c_2\right)}{2c_5^2}} + c_6 + c_7 \cdot x. \qquad (1)$$

In this function, $c_3$ is fixed at $1.490e + 01 \pm 1.04e + 00$, based on fitting the spectrum of a 1500-ppm boron solution irradiated with neutrons at the E-3 facility. Similarly, $c_5$ is fixed at $1.434e + 01 \pm 8.5e-01$, derived from fitting a spectrum measured with a $^{22}$Na point source placed at the center of the field of view. The parameters $c_1$, $c_2$, $c_4$, $c_6$, and $c_7$ are treated as free variables in the fitting process.

Herein, spectra were fitted using equation (1) in the energy range between 380 and 580 keV.

**Acknowledgment**

We would like to thank Masataka Sugiyama, CEO of SED Company, for his contributions to the development of the MPPC readout circuit for the ETCC. T.M. and S.K. also appreciate the support of all members of J-BEAM, Inc. for their involvement in this study. In addition, T.M. and S.K. acknowledge the financial support from the Fukushima Innovation Coast Framework Promotion Facility Development Subsidy, the Ministry of Economy, Trade and Industry (METI), Japan. This study was partially conducted under the Visiting Researchers Program at the Kyoto University Institute for Integrated Radiation and Nuclear Science.

**Author contributions**

T.M. and S.K. designed the experiments. T.M., S.K., and Y.S. coordinated and conducted the experiments at KURNS. T.M. wrote the draft of the manuscript, and contributed to the construction, adjustment, and operation of the ETCC as well as data analysis. T.M. and S.K. provided analysis tools used for the ETCC analysis in this manuscript. T.T., the project leader of this experiment at KURNS, provided guidance on the experimental and analytical strategies and contributed to the overall structure of the manuscript. A.T. reviewed the analysis results and provided advice on the analysis approach. All authors reviewed and commented on the manuscript.

**Data availability**

The data that support the findings of this study are available from the corresponding author upon reasonable request.

**Competing interests**

The authors declare the following competing interests: T.M. and S.K. were employed by J-BEAM, Inc. (now defunct) during the conduct of this research. The company approved the publication of this research before its dissolution. T.M., A.T., and T.T. are the inventors of the patent related to application of ETCC to environmental gamma ray measurement and its medical applications (applicant: Kyoto University, inventors: Toru Tanimori, Atsushi Takada, Tetsuya Mizumoto, and Dai Tomono, Japanese Patent Office application number P2018-520899, international application number PCT/JP2017/019936, patent currently in force). This patent is directly relevant to the techniques discussed in this manuscript. Y.S. declares no potential conflict of interest.

**Figures**

**Fig. 1: Photograph and schematic of the equipment used for measurements at the KUR E-3 facility. a** Schematic of the side of the boron solution phantom. This figure depicts a cross-sectional view of the boron solution phantom through the central axis of the cylindrical container. During the measurement, the container is positioned with the side of the polypropylene base facing downward. **b** Photograph of the ETCC measurement setup. In the figure, the yellow arrow indicate the path of the neutron beam. ETCC was placed facing upward, and the boron solution phantom was positioned on an acrylic plate placed above ETCC, aligned with the path of the neutron beam. **c** Schematic side and top views of the measurement experiment. The x–y coordinates are defined in panel **c**, with the origin at the center of the field of view. The region enclosed by the red-dashed line indicates the drift region within the TPC vessel, i.e., the detection region of TPC.

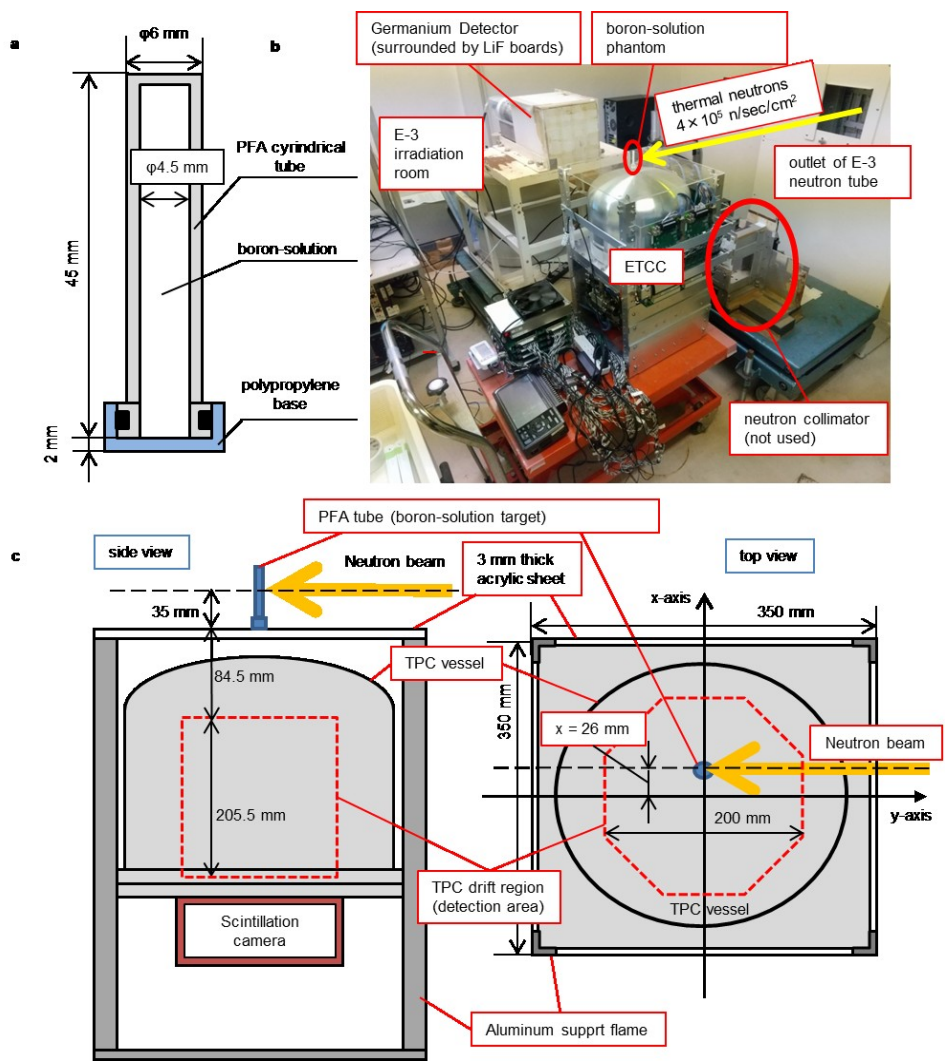

**Fig. 2: Spectra and 478-keV gamma-ray images from measurements at the KUR E-3 facility. a–d** show the results for boron solution phantoms with $^{10}$B concentrations of 0, 462, 926, and 1385 ppm, respectively. The red spectra represent residual events after applying the dE/dx, TPC fiducial volume, and 50-mm radius image cuts. The blue spectra indicate accidental coincidence events using the same cuts as the red spectra. The green spectra show the difference between the red and blue spectra. The red line in each figure represents the fit of the spectra using a composite function; the two dashed curved lines depict the Gaussian component functions for the 478 and 511-keV peaks. The bottom row presents the gamma-ray images created from events selected after the dE/dx, TPC energy, and ETCC energy cuts, processed using ML-EM**.** In each image, the yellow line and arrow indicate the central axis and direction of the neutron beam, respectively. The color scale of the images is normalized such that the maximum value of the image for **d** is 1, and the ratio of the areas within a 50 mm radius from the phantom center in the images in **a-d** was adjusted to match the ratio of the event rates within 478 keV ± 5% of the red spectrum at 478 keV in **a-d**.

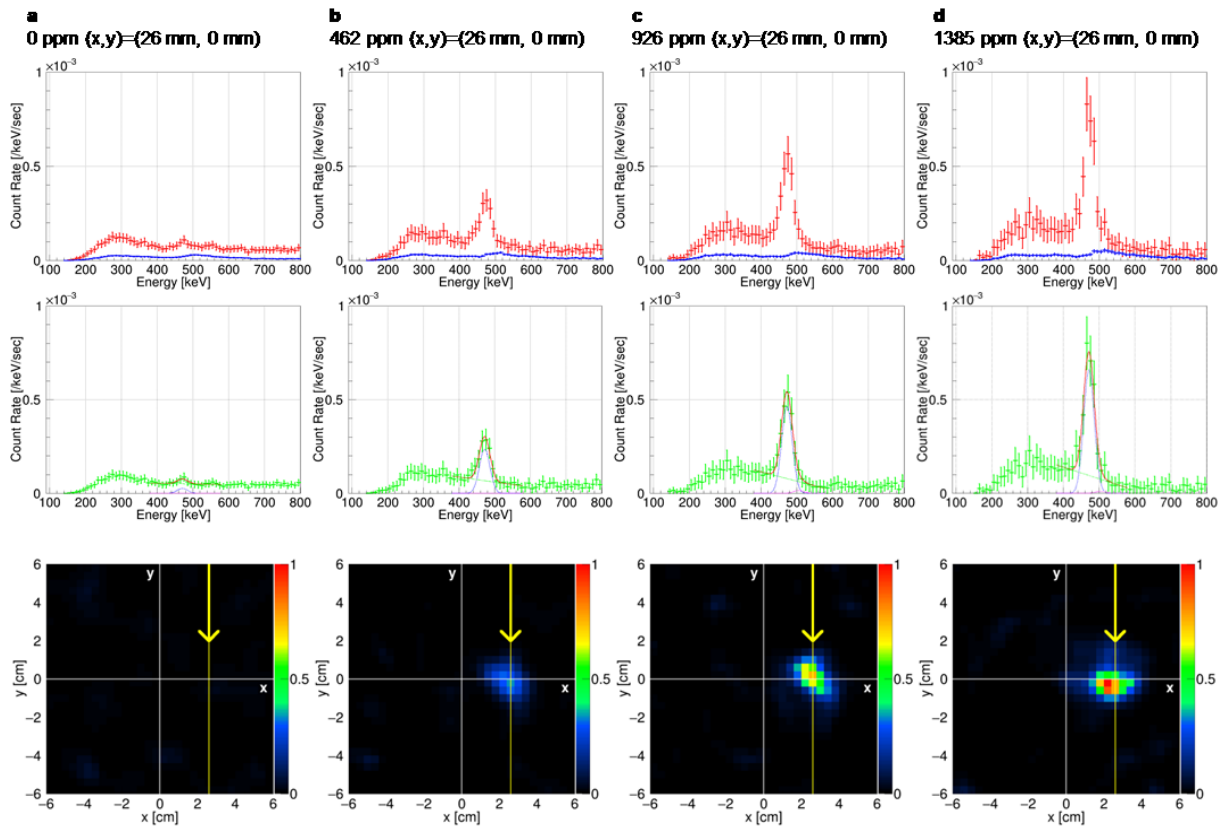

**Fig. 3: Comparisons of gamma-like spectra with accidental coincidences excluded for different boron concentrations in the phantoms and image cut radii. a** shows the superimposed spectra of four green spectra in **Fig. 2**. The red, blue, green, and magenta spectra in **a** are those for boron solution phantoms with $^{10}$B concentrations of 0, 462, 926, and 1385 ppm, respectively. The energy bin width in each energy spectrum has doubled. **b–e** show gamma-like spectra with accidental coincidences excluded for different image cut radii for boron solution phantoms with $^{10}$B concentrations of 0, 462, 926, and 1385 ppm, respectively. In **b–e**, the red, blue, and green spectra show those with image cut radii of 25, 50, and 100 mm, respectively, and the magenta spectra show those for events over the entire back-projection plane.

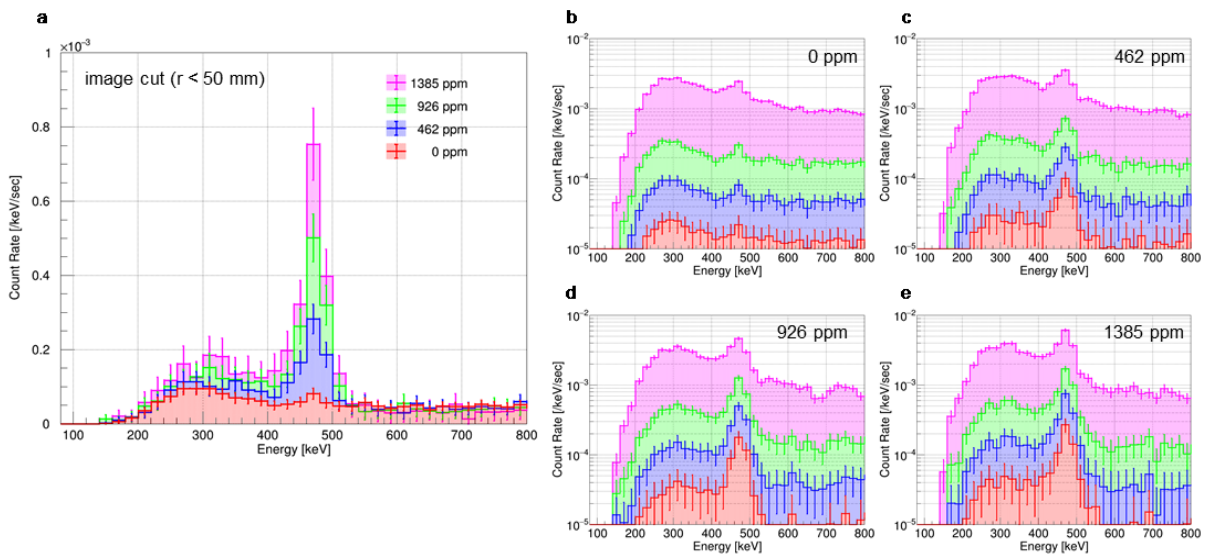

**Fig. 4: Relationship between the 478-keV gamma-ray detection rate and the $^{10}$B concentration of the solution at the KUR E-3 facility.** Magenta data points represent the 478-keV gamma-ray detection rate from the spectrum using events within a 50-mm radius in the back-projection image centered on the position of the phantom. Green data points represent the detection rate without applying the image cut, using events over the entire back-projection plane. The blue line depicts the simulated incident rate of 478-keV prompt gamma rays entering ETCC for various boron concentrations, assuming the phantom is in the same position as during the measurements. All data are normalized to the detection rate at 1385 ppm.

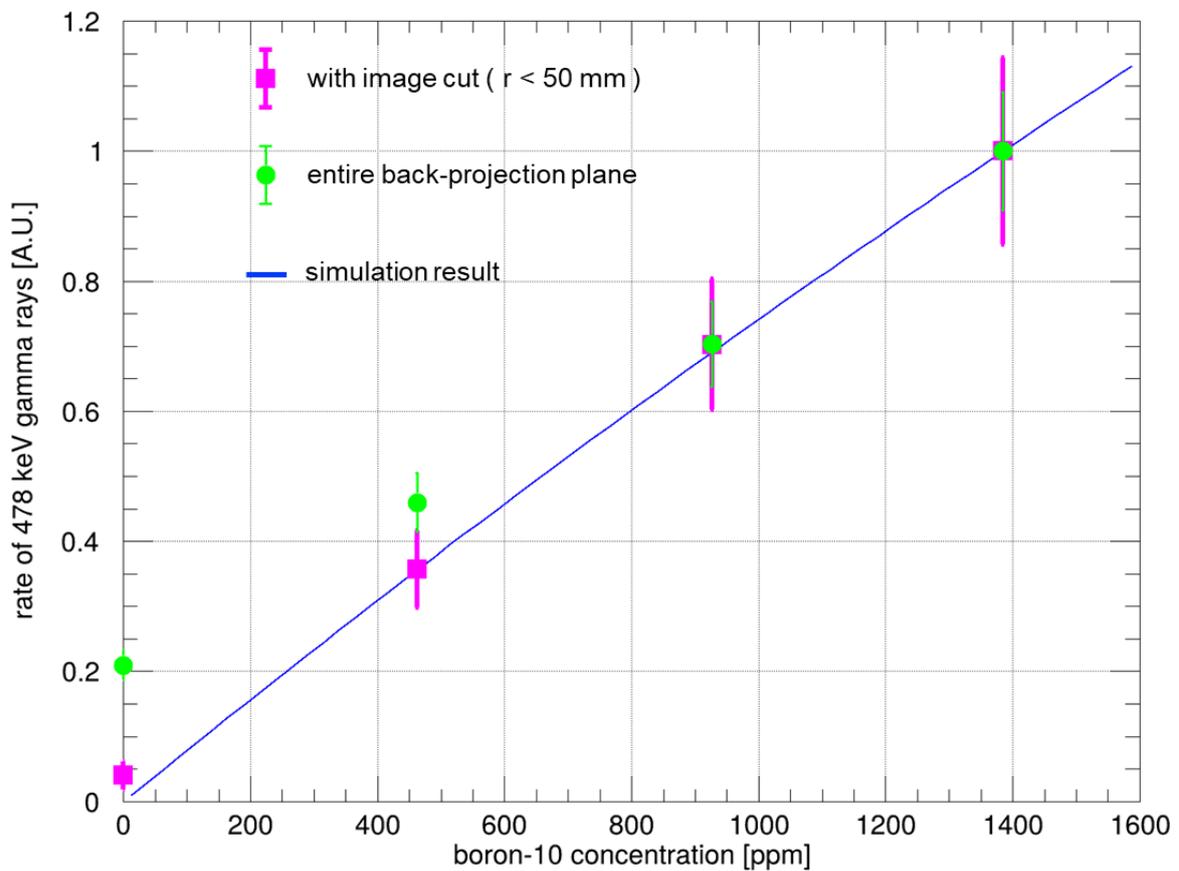

**Fig. 5: Schematic of the measurement setup.** The left panel shows the side view of the radiation room, with the KUR reactor core and HWNIF positioned to the right. This schematic shows the yz-plane at x = 0 mm. It is a cross-sectional view along a vertical plane that passes through the central axis of the ETCC, the hole in the ceiling concrete, and the center of the LiF neutron collimator. The boron-solution phantom is depicted in the form of a blue-colored square. The right panel shows the top view of the radiation room interior, with the HWNIF and KUR reactor core toward the bottom. This schematic represents the horizontal (xy) plane at the center of the neutron collimator. The boron-solution phantom is also depicted as a light-blue-colored square. The xy-coordinates are shown, where the origin is aligned with the central axis of the hole in the ceiling concrete and the central axis of the field of view of ETCC. The boron-solution phantom is positioned so that the center of its y-coordinate is aligned with y = 0.

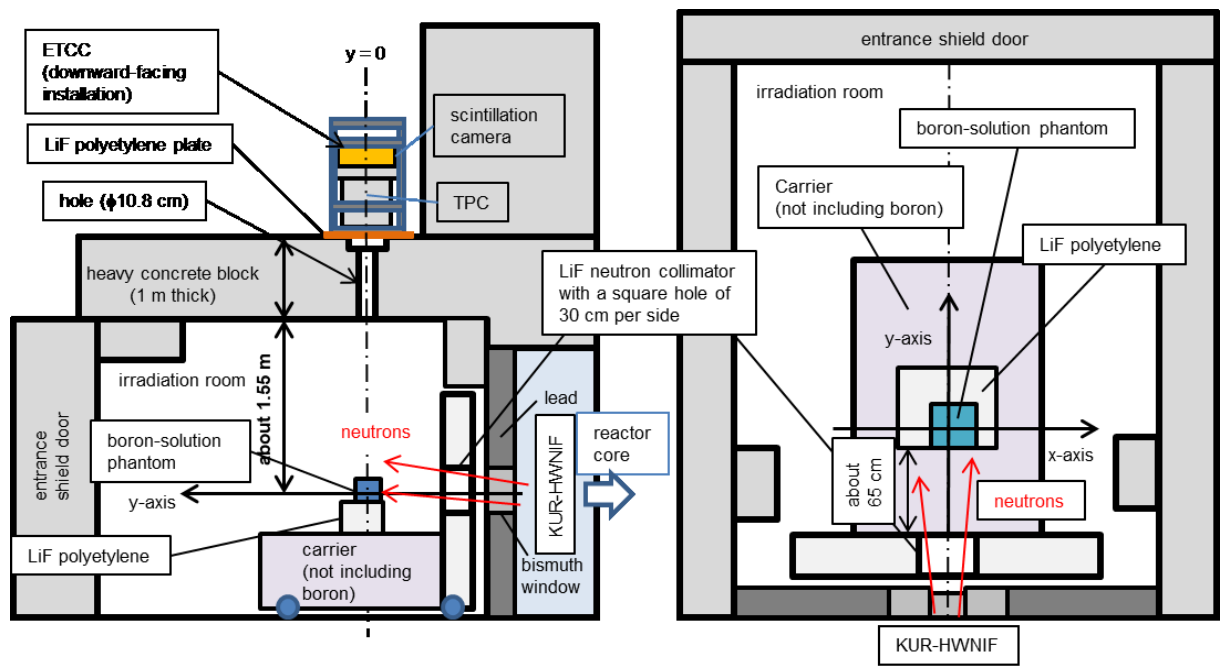

**Fig. 6: Gamma-ray images and spectra from measurements with the same phantom position and varying boron concentrations at the KUR-HWNIF.** The data corresponds to results #1 (**a**), #2 (**b**), and #3 (**c**), as detailed in Table 2. The spectra were derived using events within a 500-mm radius image cut. The red, blue, and green spectra, as well as the lines on the green spectra, are similar to those shown in **Fig. 2**. The bottom rows show ML-EM-processed gamma-ray images for energy ranges within 478 keV ± 5%. The white unfilled square in each image represents the outline of the phantom. The color scale of the ML-EM-processed images is normalized such that the maximum value of the image for **c** is 1, and the ratio of the areas within a 500-mm radius from the phantom center in the images in **a-c** was adjusted to match the ratio of the event rates within 478 keV ± 5% of the red spectrum at 478 keV in **a-c**.

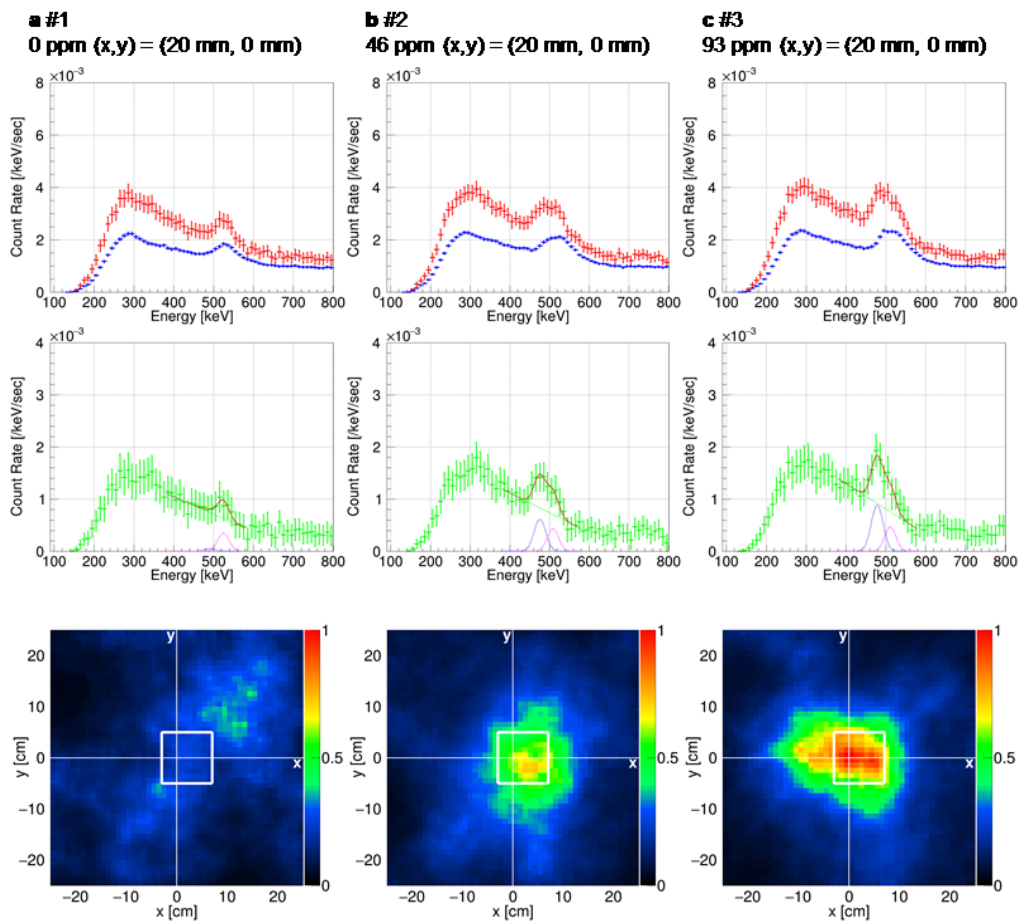

**Fig.7: Gamma-ray images and spectra from measurements with different phantom position conditions than in Fig. 6.** The data corresponds to results #4 (**a**) and #5 (**b**), as detailed in Table 2. The spectra were derived using events within a 500-mm radius image cut. The red, blue, and green spectra, as well as the lines on the green spectra, are similar to those in Fig. 2. The bottom rows show ML-EM-processed gamma-ray images for energy ranges within 478 keV ± 5% for **b** and the left image of **a** and those within 511 keV$^{+5\%}_{-0\%}$ for the right image of **a**. The white unfilled square in each image outlines the phantom. The color scale of the two images of **a** is normalized such that the maximum value of the left image is 1. The ratio of the areas within a 500 mm radius from the 230-ppm phantom center in the left image to that in the right image was adjusted so that it matches the ratio of the event rates within 478 keV ± 5% of the red spectrum to twice the event rates within 511 keV$^{+5\%}_{-0\%}$ of the same spectrum. The color scale of the ML-EM-processed image of **b** is normalized from 0 to 1.

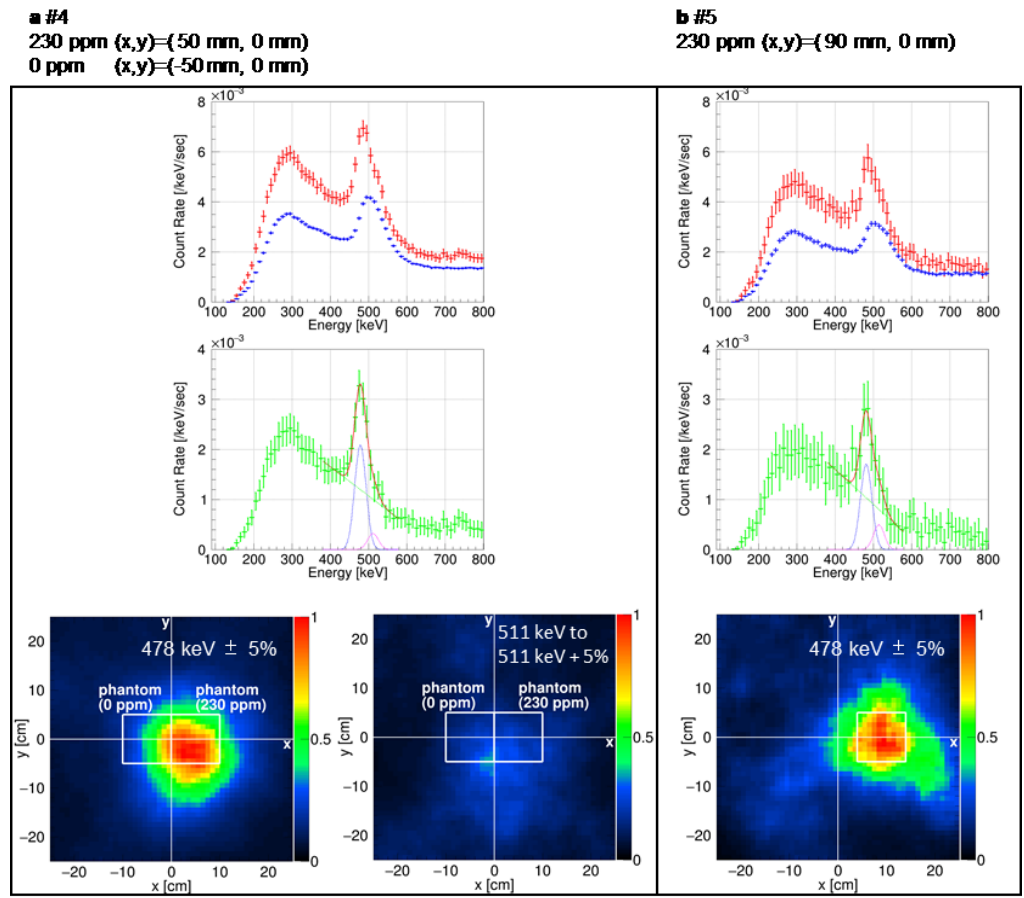

**Fig. 8: Relationship between the 478-keV gamma-ray detection rate and the $^{10}$B concentration in the solution at the KUR-HWNIF.** The magenta and green points as well as the blue curve, are similar to those in **Fig. 4**. The green data points correspond to an image cut radius of 500 mm. The blue line shows the simulated incoming rate of 478-keV prompt gamma rays into ETCC for different boron concentrations, assuming the position of the phantom is the same as in measurements #1–#3, as detailed in Table 2. The simulations, conducted with PHITS, considered the phantom, air, ceiling concrete, and drilled holes. All data are normalized relative to the 93 ppm. As measurement #4 had different irradiation conditions compared with measurements #1–#3, the 230-ppm data points were adjusted using a ratio based on the simulation results for measurement #4.

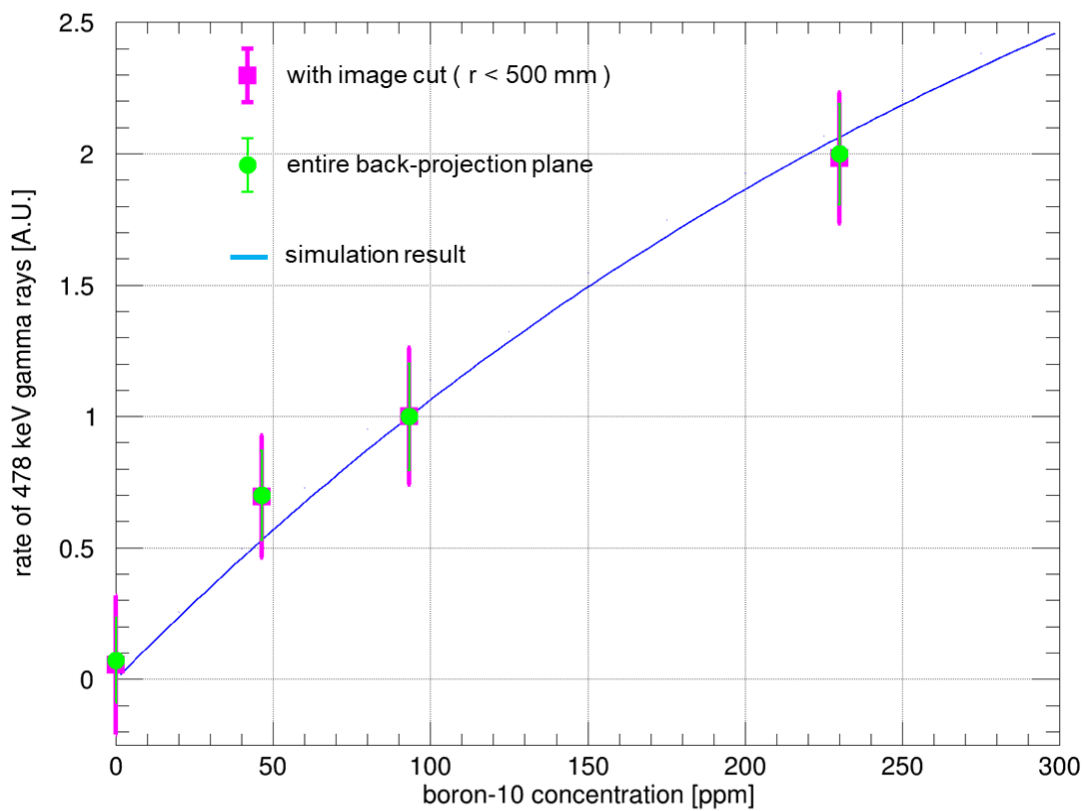

**Tables**

**Table 1. Measurement times and results for the averaged signal rates, dead time, data acquisition rates, and 478-keV gamma-ray detection rates for each $^{10}$B concentration of the solution measured with ETCC at the E-3 facility.** All rates are averaged over the entire measurement time. The 478-keV gamma-ray detection rate represents the event rate selected by event selection criteria 1–4 in the "ETCC data analysis method" subsection of the "methods" section.

| $^{10}$B concentration [ppm] | Measurement time [sec] | TPC signal rate [Hz] | Scintillation camera signal rate [Hz] | TPC data acquisition rate [Hz] | Dead time [%] | 478-keV gamma-ray detection rate [Hz] |
|---|---|---|---|---|---|---|
| 0 | 23943.56 | 820.21 | 6031.29 | 32.33 | 0.14 | 0.14 |
| 462 | 9868.21 | 810.19 | 6057.44 | 31.96 | 0.16 | 0.18 |
| 926 | 6460.05 | 775.06 | 5904.94 | 30.38 | 0.13 | 0.23 |
| 1385 | 4179.21 | 754.44 | 6042.31 | 30.42 | 0.13 | 0.28 |

**Table 2. Measurement conditions and results for the signal rate, data acquisition rate, and dead time for each measurement conducted in the KUR-HWNIF irradiation room.** All rates have the same meaning as those in **Table 1**.

| Number | Target center position (x,y) [mm] | $^{10}$B concentration [ppm] | Measurement time [sec] | TPC signal rate [Hz] | Scintillation camera signal rate [Hz] | TPC data acquisition rate [Hz] | dead time [%] | 478-keV gamma-ray detection rate [Hz] |
|---|---|---|---|---|---|---|---|---|
| 1 | (20, 0) | 0 | 3527.63 | 6835.91 | 64857.25 | 590.98 | 2.37 | 2.23 |
| 2 | (20, 0) | 46 | 4705.87 | 6762.44 | 66237.16 | 602.6 | 2.42 | 2.67 |
| 3 | (20, 0) | 93 | 4261.69 | 6584.74 | 67733.8 | 615.9 | 2.48 | 2.99 |
| 4 | (50, 0) / (−50, 0) | 230 / 0 | 7335.40 | 4772.86 | 80710.31 | 760.83 | 3.04 | 4.46 |
| 5 | (90, 0) | 230 | 2081.3 | 4037.12 | 68364.9 | 599.15 | 2.38 | 3.59 |